\documentclass[preprint,showkeys,secnumarabic,showpacs,amsmath,amssymb]{revtex4}
\usepackage{graphicx}

\begin{document}

\title{On Dynamics of Brans--Dicke Theory of Gravitation}

\author{Hossein Farajollahi}
 \email{hosseinf@guilan.ac.ir}
\affiliation{Department of Physics, University of Guilan, Rasht, Iran}

\author{Mehrdad Farhoudi}
 \email{m-farhoudi@sbu.ac.ir}

 \author{Hossein Shojaie}
 \email{h-shojaie@sbu.ac.ir}

\affiliation{Department of Physics, Shahid Beheshti University, G.C., Evin, Tehran 19839, Iran}

\date{\today}

\begin{abstract}
We study longstanding problem of cosmological clock in the context
of Brans--Dicke theory of gravitation. We present the Hamiltonian
formulation of the theory for a class of spatially homogenous
cosmological models. Then, we show that formulation of the
Brans--Dicke theory in the Einstein frame allows how an
identification of an appropriate cosmological time variable, as a
function of the scalar field in the theory, can be emerged in
quantum cosmology. The classical and quantum results are applied
to the Friedmann-–Robertson-–Walker cosmological models.
\end{abstract}



\maketitle

\section{Introduction}

Identifying a suitable cosmological clock with respect to which
the dynamics of the Universe can be measured is one of the most
fundamental problems in quantum cosmology. The conventional
Wheeler--DeWitt~({\bf WD}) formulation gives a time--independent
quantum theory, and consequently, observers experience the passage
of time. Although the problem manifests itself in quantum level,
it is originated from classical theories of gravity, in which  the
theory is invariant under time re--parametrization, see, e.g.,
Refs.~\cite{Benedetti,Kiefer,Gambini} and references therein.

In the context of general relativity, a lot of efforts have been
directed to solve the problem via re--interpretation of time
notion, or introducing a material time variable emerged from the
geometry or matter of four dimensional space--time, and or
introduction of any other kind of geometrical
objects~\cite{Isham,Kuchar,Wald,Halliwell,Torre,Higuchi,Feinberg,
Smolin,Gaioli,Rovelli,Rovelli1,Rovelli2,Rovelli3,Rovelli4}. In
particular, introduction of a scalar field in the time
re--parametrization invariant theories, e.g. general relativity,
to present a cosmological clock has been
investigated~\cite{Hajicek,Hajicek1,capozziello1996,Isham1,Thiemann,Shestakova1,farajollahi,farajollahi1}.
Unfortunately, there still does~not exist a completely
satisfactory solution to this fundamental issue.

On the other hand, physicists have long entertained the
scalar--tensor theories of gravitation as alternative ones and the
most popular rivals to the general relativity, in which the nature
simplest imaginable phenomenon, namely a scalar field, plays the
major role. The scalar--tensor theory, developed originally by
Jordan, began by embedding a four dimensional curved manifold in
five dimensional flat space--time~\cite{Jordan}.

The Brans--Dicke gravitational theory is the well--known example
of the scalar--tensor theories, in which the gravitational
interaction, besides the usual metric of general relativity, is
mediated by a scalar field, which has the physical effect of
changing the effective gravitational constant from place to place.
According to the Brans--Dicke assumption~\cite{Brans}, the scalar
field decoupled from the matter part of the Lagrangian in order to
save the weak equivalence principle being violated. However, this
assumption hardly seems to be supported by any example of more
fundamental theories and alternatively, it has been shown that the
size of a compactified internal space can behave as a four
dimensional scalar field of the nature of the Brans--Dicke model,
with a parameter determined uniquely in terms of the
dimensionality of space--time~\cite{fujii}.

In the cosmological context, scalar fields have been supposed to
cause an accelerating expanded universe~\cite{Guth}, to solve the
horizon problem and to give a hypothetical reason for the
non--vanishing cosmological constant. Massless, or long--ranged,
scalar fields, in this context, are known as inflatons and
massive, or short--ranged, scalar fields are proposed to use, for
example, Higgs--like fields~\cite{Cota}. In this work, we purpose
to identify another role for the scalar field in the Brans--Dicke
theory in order to solve the problem of time in quantum cosmology.
Since the usual time coordinate in four dimensional space--time
is~not a suitable candidate to play the role of cosmological time,
one may speculate that the dynamical Brans--Dicke scalar field --
decoupled from matter in the Lagrangian or originated from higher
dimensional space -- is eligible to perform such a role similar to
the role of Newtonian external time parameter in the classical
mechanics.

The non--minimal coupling term in the scalar--tensor theory is
equivalent to the presence of conformal transformations. That is,
one can change the form of this coupling term by applying these
transformations. Among these transformations, the ones which can
transform the coupling term into a constant are more attractive,
for they can be employed to switch between the Jordan and Einstein
frames, which are the most discussed conformal frames ~\cite{faraon1999,martin2010}. Some
aspects of the cosmological conformal equivalence between these
two frames have also been investigated~\cite{capozziello1996}.
Nonetheless, the physics is not invariant under conformal
transformations, except in the weak gravitational field limit, and
one should choose one of them as the physical frame. As the
physical frame labels the frame we live in, it can be selected
upon physical grounds consistent with principles and observations.

In this manuscript, in section two, we start with the description
of the Brans--Dicke theory and present it in the Einstein frame.
The advantage of Einstein frame over the Jordan frame is that the
scalar field contribution to the Brans--Dicke Lagrangian density
is identical to one used in the vacuum general relativity, where
it gives the possibility of emerging a cosmological time variable
in the theory as we will discuss in this work. In section three,
the Hamiltonian formulation of the theory is presented. The
formulation is given for spatially homogenous cosmological models
which provide good prototypes for many theoretical models in
gravity and cosmology. Then, a cosmological time variable is
introduced, as a function of the scalar field in the theory, in
which the dynamics of the metric functions can be performed. In
section four, as an example, the classical and quantum dynamics of
the Friedmann--Robertson--Walker~({\bf FRW}) models are studied.
Finally, section five presents conclusions and remarks drawn from
this work.

\section{Brans--Dicke Theory of Gravitation in Einstein Frame}

The Brans--Dicke theory of gravitation, based on the Mach
principles, is presented by a general action for a real scalar
field living in a four dimensional curved space--time with the
(physical) Lagrangian density given by
\begin{equation}
{\cal L}_{\rm BD}=\sqrt{-g}\left(\varphi
R-\frac{\omega}{\varphi}g^{\mu\nu}\partial_\mu \varphi
\partial_\nu\varphi\right)
\label{BDlagrang},
\end{equation}
where the lower Greek indices run from zero to three, $R$ is the
Ricci curvature scalar formed from the metric $g_{\mu\nu}$. The
positive value $\varphi(x^\rho)$ is called Jordan's scalar field
and the dimensionless constant $\omega$ is the only parameter of
theory. We assume that there is no other matter field in the
theory. The $\varphi R$ term is a non--minimal coupling term, in
which $G^{-1}$ in the Einstein--Hilbert term is replaced with
$\varphi$. This provides that, as long as the scalar field varies
slowly, the Universe is assumed to feel an effective spatially
uniform gravitational constant $G_{\rm eff}=1/\varphi $\ which
depends only on the cosmic time. The second term is proportional
to the kinetic term of the scalar field. To remove the singular
behavior of $\varphi^{-1}$ from this term, one can replace it by
introducing a new real scalar field, as
$\phi=\sqrt{8|\omega|\varphi}$\,, that converts~(\ref{BDlagrang})
to
\begin{equation}
{\cal L}_{\rm BD}=\sqrt{-g}\left(\frac{1}{8|\omega|}\phi^2
R-\frac{1}{2}\epsilon g^{\mu\nu}\partial_\mu\phi
\partial_\nu\phi\right)
\label{BDlagrang1},
\end{equation}
where $\epsilon=\pm1$ keeps the sign of $\omega$. It is worth
noting that when $\omega<0$, the scalar field can be interpreted
as a ghost~\cite{fujii}.

The Lagrangian density~(\ref{BDlagrang1}) has been written in the
Jordan frame as a physical frame, however we intend to proceed in
the Einstein frame. This can be performed by employing a conformal
transformation such that it transforms the coefficient of $R$ to a
constant, where the causality structure does~not change.
Mathematically, the conformal transformation
$\widehat{g}_{\mu\nu}=\Omega^2(x^\rho)g_{\mu\nu}$
transforms~(\ref{BDlagrang1}) to
\begin{equation}
{\cal L}_{\rm
BD}=\sqrt{-\widehat{g}}\Omega^{-2}\left\{\frac{1}{8|\omega|}\phi^2\left[\widehat{R}
+6\,\widehat{\Box}\ln\Omega+6\,\widehat{g}^{\mu\nu}\partial_\mu(\ln\Omega)\partial_\nu(\ln\Omega)\right]
-\frac{1}{2}\epsilon\,\widehat{g}^{\mu\nu}\partial_\mu\phi\partial_\nu\phi\right\}
\label{BDlagrang2}.
\end{equation}
Choosing $\Omega=\phi/\sqrt{8|\omega|}$ and eliminating the second
term of the Lagrangian density by integrating by parts, it yields
\begin{equation}
{\cal L}_{\rm BD}=\sqrt{-\widehat{g}}\left(\widehat{R}
-\frac{1}{2}\lambda\,\widehat{g}^{\mu\nu}\partial_\mu\phi\partial_\nu\phi\right)
\label{BDlagrang3},
\end{equation}
where
\begin{equation}
\lambda\equiv\Omega^{-2}\left[\epsilon
+12\left(\frac{d\Omega}{d\phi}
\right)^2\right]=\frac{8}{\phi^2}\left(\epsilon\,|\omega|+\frac{3}{2}\right)\label{lambda}.
\end{equation}
For $\phi$ to be a normal field instead of a ghost, $\lambda$ is
positive when $\omega>-3/2$. Actually, this restriction depends on
the dimension of space--time $D$, namely
$\omega>-(D-1)/(D-2)$~\cite{BKM04,DDB07}.

Now, one can introduce another new scalar field, by
$\partial_\mu\psi=\sqrt{\lambda}\,\partial_\mu\phi $\,, and
finally rewrites~(\ref{BDlagrang3}) as
\begin{equation}
{\cal L}_{\rm BD}=\sqrt{-\widehat{g}}\left(\widehat{R}
-\frac{1}{2}\widehat{g}^{\mu\nu}\partial_\mu\psi\partial_\nu\psi\right)
\label{BDfinal},
\end{equation}
which is obviously the Lagrangian density in the Einstein frame
with a canonical kinetic term.

The Lagrangian density~(\ref{BDfinal}) is actually the well--known
Lagrangian of vacuum general relativity plus contribution from a
scalar field. However, one usually lacks physical justifications
to {\it priori} introduce such a kinetic term when working with
vacuum general relativity, although via the induced-–matter
theory~\cite{wessonBook1999,wessonBook2006}, one may be able to
get such an effective induced kinetic term. Nevertheless, through
this work and by introducing a cosmological time, we will also
offer a physical base to justify starting with the Lagrangian
density~(\ref{BDfinal}) right at the beginning.

\section{Hamiltonian Formulation and Cosmological Time}

While in the present work, for its relative simplicity, a
Lagrangian formulation of the theory is given, we are more
interested in the Hamiltonian version of the theory where the
absence of time displays itself in a freezing Hamiltonian. For
this purpose, we consider the Hamiltonian formulation of the
theory.

Suppose that the four dimensional space-time manifold can be
foliated by homogeneous, but generally anisotropic, spatial
hypersurfaces $\Sigma_t$ of constant time t. Then, the associated
space--time metric can have the form
\begin{equation}
ds^2=-N^2(t)dt^2+\widehat{g}_{ij}dx^idx^j
    =-N^2(t)dt^2+a^2(t)\gamma_{ij}dx^idx^j\label{homometric},
\end{equation}
where the lower Latin indices run from one to three, $N(t)$ is a
lapse function and $\gamma_{ij}$ is the time--independent spatial
3--metric on the hypersurfaces $\Sigma_t$. For convenience, we
assume that the time axis is always normal to the hypersurfaces of
homogeneity $\Sigma_t$. We also restrict the scalar field
(dilaton) to be spatially homogeneous.

From Lagrangian~(\ref{BDfinal}) and metric~(\ref{homometric}), the
canonical momentum associated with the dynamical variable $\psi$
is
\begin{equation}
\pi_\psi=\sqrt{{}^{(3)}\widehat{g}}\,N^{-1}\dot{\psi}\label{momentumconst},
\end{equation}
where ${}^{(3)}\widehat{g}=a^6(t)\gamma$ is the determinant of
3--metric on $\Sigma_t$ and $\gamma$ is the determinant of
$\gamma_{ij}$. Following the Dirac procedure~\cite{Dirac}, the
total Hamiltonian obtained from the action, by a Legendre
transformation, is
\begin{equation}
H_{\rm tot}=\int_{\Sigma_t} N\left({\cal
H}+\frac{\pi_\psi^{2}}{2\sqrt{{}^{(3)}\widehat{g}}} \right)d^3x
\label{totalHamiltonian}.
\end{equation}
The variable $N$ plays the role of a Lagrange multiplier and
${\cal H}$ is the super--Hamiltonian, generally, given by
\begin{align}
{\cal H}(t, x^i;\widehat{g}_{ij},\pi^{ij})={\cal G}_{mnkl}(t,
x^i)\,\pi^{mn}(t, x^i)\,\pi^{kl}(t,
x^i)-\sqrt{{}^{(3)}\widehat{g}}\, R(t,
x^i,\widehat{g}_{ij})\label{SuperHam},
\end{align}
in which
\begin{equation}
{\cal G}_{mnkl}(t, x^i)=
\frac{1}{2\sqrt{{}^{(3)}\widehat{g}}}\Big[\widehat{g}_{mk}(t,
x^i)\,\widehat{g}_{nl}(t, x^i) +\widehat{g}_{nk}(t,
x^i)\,\widehat{g}_{ml}(t, x^i)-\widehat{g}_{mn}(t,
x^i)\,\widehat{g}_{kl}(t, x^i)\Big]\label{SuperMet}
\end{equation}
is the DeWitt supermetric on the space of 3--metrics and
$\pi^{ij}$ is the conjugate momentum of $\widehat{g}_{ij}$ . We
thus have a new first class Hamiltonian constraint,
\begin{equation}
{\cal H}+\frac{\pi_\psi^{2}}{2\sqrt{{}^{(3)}\widehat{g}}}\approx 0\label{HamCon},
\end{equation}
instead of ${\cal H}\approx 0$. By the non--degenerate character
of the metric, the new Hamiltonian constraint can be redefined by
rescaling $N$ as $N_\ast=N/a^3$. Namely
\begin{equation}
{\cal H}_\ast=a^3{\cal H}+\frac{\pi_\psi^2}{2\sqrt{\gamma}}\approx
0 \label{totalHamiltonian1},
\end{equation}
where the total Hamiltonian then becomes
\begin{equation}
H_{\rm tot}=\int_{\Sigma_t}N_\ast{\cal H}_\ast
d^3x\label{newtotalHamiltonian}.
\end{equation}

In the Hamiltonian formulation of the theory, one can perform a
canonical transformation from ($\psi, \pi_\psi$) to ($T, \Pi_T$)
that preserves the form of equations of motion. Choosing the new
canonical coordinate as
\begin{equation}
T(t)=\frac{\psi}{\pi_\psi}\gamma\label{Time},
\end{equation}
then by calculating the Dirac bracket relation, one finds that
\begin{equation}
\Pi_T=\frac{\pi_\psi^2}{2\gamma}\, .
\end{equation}

An analogy with unimodular general
relativity~\cite{Unruh,Henneaux,Unruh1} arises an anticipation
that this new variable $T(t)$ may play the role of a cosmological
time variable. This hope is fulfilled. Actually, the equation of
motion derived from~(\ref{Time}) implies that
\begin{equation}
\frac{dT}{dt}=[T,H_{\rm
tot}]=\int_{\Sigma_t}N_\ast\sqrt{\gamma}d^3x\label{cosmictime1},
\end{equation}
and hence $N_\ast\sqrt{\gamma}$ presents the density of $\dot{T}$
over the hypersurface. By integrating, one obtains
\begin{equation}
T(t)=\int N_\ast {}^{(3)}\!V\!_{_{\rm
proper}}dt\label{cosmictime2}.
\end{equation}
Therefore, $T(t)$ is just the 4--volume between the hypersurfaces
$\Sigma_{t_0}$ and $\Sigma_t$ with $t_0$ as an initial time, and
equation~(\ref{cosmictime1}) shows that the rate of change of this
time variable is necessarily a positive value.

In the Einstein frame formulation of the Brans--Dicke cosmology,
$T(t)$ coincides classically with the cosmological time parameter
which arises both in the unimodular general
relativity~\cite{Unruh,Henneaux,Unruh1,Hossein} and in Sorkin's
sum over histories approach~\cite{Sorkin}. Note that, $T(t)$ is a
monotonically increasing function along any future directed
time--like curve, and thus can indeed be used to parameterize this
trajectory.

The gauge invariant quantity $T(t)$ is~not a Dirac observable, for
it does not commute with ${\cal H}_\ast$, but its conjugate
momentum, $\Pi_T$, is a Dirac observable. Besides, $T(t)$ can be
used as a clock. Indeed, if we label the spatial hypersurfaces by
the cosmological time $T(t)$ instead of the coordinate time $t$,
and foliate the space--time by $T(t)$ and $\Sigma_T$ rather than
$t$ and $\Sigma_t$, then any geometric, i.e. generally covariant,
quantity defined on $\Sigma_T$ has vanishing Poisson bracket with
the integrated Hamiltonian constraint.

Note that, in the formulation of the theory in terms of the old
canonical variables, ($\psi, \pi_\psi$), in quantizing the theory
one obtains a Klein--Gordon like equation and needs to perform  a
decomposition of the solution in order to get a
Schr\"{o}dinger--like equation. However, in terms of the new
canonical variables, ($T, \Pi_T$), we directly  acquire a
Schr\"{o}dinger--like equation. Indeed, in the quantum theory, the
momentum variable $\Pi_T$ is represented by the operator
$-i\partial/\partial T$ and the WD equation of the Hamiltonian
constraint~(\ref{totalHamiltonian1}) takes the form of a
Schr\"{o}dinger equation describing the evolution of cosmological
wave functions, e.g. $\Psi$, with respect to the cosmological time
parameter, $T(t)$, as
\begin{equation}
i\frac{\partial\Psi}{\partial T}=\frac{a^3}{\sqrt\gamma}\,{\cal
H}\Psi .
\end{equation}

In order to realize properties and effects of such a cosmological
time, we probe the dynamical behaviour of the universe in the FRW
models with respect to this clock in the next section.

\section{The FRW Models}

As an example, we begin with the line element for the FRW models
in the spherical coordinates, namely metric~(\ref{homometric}) in
which $a(t)$ is the cosmic scale factor -- that determines the
radius of universe -- and $\gamma_{ij}$ is the time--independent
metric of the three--dimensional maximally symmetric spatial
sections
\begin{equation}
\gamma_{ij}dx^idx^j = \frac{dr^2}{1 - kr^2} + r^2 (d\theta^2 +
{\sin^2}\theta d\phi^2),
\end{equation}
with the constant curvature scalar $^{(3)}R(\gamma_{ij})=6k$.

The Brans--Dicke action in the Einstein frame for the FRW models
with a minimally coupled massless scalar field to gravity is given
by
\begin{equation}\label{action-frw}
S[g_{\mu\nu} , \psi] =\int\left[6\left(-\frac{a{\dot
a}^2}{N}+kNa\right) + \frac{a^3}{2N}
{\dot\psi}^2\right]\sqrt{\gamma}\,d^4x\,.
\end{equation}
Hence, the total Hamiltonian is
\begin{equation}
H_{\rm tot}=\int_{\Sigma_t}
N\left(-\frac{p_a^2}{24a\sqrt{\gamma}}- 6ka\sqrt{\gamma} +
\frac{\pi_\psi^2}{2a^3\sqrt{\gamma}}\right)d^3x\label{Hamiltonian}.
\end{equation}
Now, one can write the total Hamiltonian as
\begin{equation}
H_{\rm tot}=\int_{\Sigma_t}
N_\ast\left(-\frac{a^2p_a^2}{24\sqrt{\gamma}}- 6ka^4\sqrt{\gamma}+
\Pi_T\sqrt{\gamma}\right)d^3x\label{finalHamiltonian},
\end{equation}
where again we apply $N_\ast=N/a^3$ as a rescaling. Following the
procedure in the previous section, the first class constraint can be
redefined as
\begin{equation}
0 \approx {\cal H}_\ast =-\frac{a^2p_a^2}{24\sqrt{\gamma}} - 6ka^4\sqrt{\gamma} +\Pi_T\sqrt{\gamma}\label{hamilconst}.
\end{equation}
The classical equations can be obtained from
(\ref{finalHamiltonian}) as
\begin{equation}
  \dot{a}=\frac{\partial (N_\ast{\cal H}_\ast)}{\partial p_a}=-\frac{N_\ast a^2p_a}{12\sqrt{\gamma}},
\end{equation}
\begin{equation}
  \dot{p}_a=-\frac{\partial (N_\ast{\cal H}_\ast)}{\partial a}=N_\ast\left(\frac{ap_a^2}{12\sqrt{\gamma}}+24ka^3\sqrt{\gamma}\right),
\end{equation}
\begin{equation}
  \dot{T}_{\rm pd}=\frac{\partial (N_\ast{\cal H}_\ast)}{\partial \Pi_T}=N_\ast \sqrt{\gamma},
\end{equation}
and
\begin{equation}
  \dot{\Pi}_T=-\frac{\partial (N_\ast{\cal H}_\ast)}{\partial
  T}=0,
\label{pidot}
\end{equation}
where $\dot{T}_{\rm pd}$ is the proper density of cosmological
time. Equation~(\ref{pidot}) shows that $\Pi_T$ is a constant of
motion. By relation~(\ref{momentumconst}), one gets $\dot{T}_{\rm
pd}=\gamma\dot{\psi}$ which simply shows that the dynamics of the
cosmological time variable, $T$, is proportional to the dynamics
of the scalar field.

We are more interested to employ the cosmological time, $T(t)$,
and find the behavior of the isotropic variables $a$ and $p_a$
with respect to it. Thus, we find
\begin{equation}
  a'=-\frac{a^2p_a}{12\gamma}\label{aprime}
\end{equation}
and
\begin{equation}
  p'_a=\frac{ap_a^2}{12\gamma}+24ka^3\label{pprime}
\end{equation}
where the prime denotes derivative with respect to $T$. In the
following, we investigate the behavior of scale factor with
respect to the cosmological time $T$ for various curvatures.
\begin{description}
\item[Case] ${\bf k=0}$:

Solving the last two equations~(\ref{aprime}) and~(\ref{pprime}),
assuming $k=0$, gives
\begin{equation}
    a(T)=a_0e^{-C\,T}
\end{equation}
and
\begin{equation}
    p_a(T)=p_{a_0}e^{C\,T}
\end{equation}
where $a_0=a|_{T=0}$, $p_{a_0}=p_a|_{T=0}$ and $C\equiv
a(T)p_a(T)/12$ is a constant. A negative $C$ with positive $a_0$
provides an accelerating expanded universe, with a positive
constant Hubble, $H=-C$. In this case, universe expands
exponentially according to the cosmological time variable, and
naturally accelerating without a beginning singularity. Indeed,
the monotonic dependence of the dynamical clock $T$ with respect
to the scale factor shows that it can be employed as a
cosmological time for the gravitational dynamics.

\item[Case] ${\bf k=1}$:

The nontrivial solution of the above coupled non linear
differential equations~(\ref{aprime}) and (\ref{pprime}), for the
scale factor gives
\begin{equation}
  a(T)=\frac{\sqrt{2\sqrt{c_1}c_2}\, e^{\sqrt{c_1}T}}{\gamma^{1/4}(1+c_2^2
  e^{4\sqrt{c_1}T})^{1/2}},
\end{equation}
where $c_1$ and $c_2$ are positive constants of integrations. This
solution shows that universe has no singularity at all and for
small enough $T$ the scale factor also becomes very small. The
solution goes to zero when $T$ goes to minus infinity. Besides,
universe shrinks to a big crunch as $T$ goes to infinity while it
reaches a maximum size during its history.

\item[Case] ${\bf k=-1}$:

In this case the solution is
\begin{equation}
  a(T)=\frac{\sqrt{2\sqrt{c_3}c_4}\, e^{\sqrt{c_3}T}}{\gamma^{1/4}(1-c_4^2
  e^{4\sqrt{c_3}T})^{1/2}},
\end{equation}
where again $c$'s are positive constants of integrations. This
solution shows that the scale factor goes to infinity for some
$T$, depends on the constants of integrations, and becomes
imaginary beyond it. Such a scale factor does~not have a well
physical interpretation.
\end{description}

When the system is canonically quantized, the associated WD
equation, i.e. the Schr\"{o}dinger--like equation, describes how
the wave function of universe evolves with the cosmological time
variable. From equation~(\ref{hamilconst}), we have
\begin{equation}
-\partial_a^2 \Psi+144k\gamma a^2\Psi
+i\frac{24\gamma}{a^2}\partial_T\Psi=0,
\end{equation}
where for wave function $\Psi(a,T)$ as
\begin{equation}
\Psi_E(a,T)=\Psi(a)e^{iET},
\end{equation}
$\Psi(a)$ satisfies
\begin{equation}
-\frac{a^2}{24\gamma}\partial_a^2\Psi(a)+(6k a^4)\Psi(a)=E\Psi(a)
\label{spatial-sch-eq} .
\end{equation}

In order to have a self--adjoint
Hamiltonian, the inner product between
wave functions must be
\begin{equation}
<\Phi,\Psi >(T)=\int_0^{\infty}a^{-2}\Phi^*\Psi da,
\end{equation}
and solutions must satisfy the convenient boundary conditions
\begin{equation}
\Psi(0, T)=0\qquad\text{or}\qquad \frac{\partial\Psi(a,
T)}{\partial a}{\Bigr |}_{a=0}=0\label{bdry},
\end{equation}
in the domain of Hamiltonian operator~\cite{lemos96,alvarenga98}.

In the following, we find cosmological wave functions for various
curvatures.
\begin{description}
\item[Case] ${\bf k=0}$:

For spatially flat space--time, equation~(\ref{spatial-sch-eq}) is
an Euler--type equation
\begin{equation}
\partial_a^2\Psi(a)+24\gamma a^{-2} E\Psi(a)=0\label{spatialscheqk=0},
\end{equation}
and its general solution is
\begin{equation}
\Psi_E(a, T)=e^{iET}\sqrt{a}\left(c_5a^{\frac{\sqrt{1-96\gamma
E}}{2}}+c_6a^{-\frac{\sqrt{1-96\gamma E}}{2}}\right)\label{solfor
k=0},
\end{equation}
where $c$'s are constants of integrations. Obviously, the above
solution is~not square integrable and in order to obtain a
possible physical solution, one should construct wave packets as
\begin{equation}
\Psi(a, T)=\int_0^{\infty}A(E)\Psi_E(a, T)dE.
\end{equation}
However, even with the above wave packets, still the ill--behavior
of solution~(\ref{solfor k=0}) prevents one to get finite--norm
states by superposing them.

\item[Case] ${\bf k=1}$:

The solution for positive curvature is
\begin{equation}
\Psi_E(a, T)=e^{iET}\sqrt{a}\left[c_7K_\nu(6\gamma
a^2)+c_8I_\nu(6\gamma a^2)\right],
\end{equation}
where again $c$'s are constants of integrations, $K_\nu$ and
$I_\nu$ are the modified Bessel functions and
$\nu=\sqrt{1-96\gamma E}/4$. Since $I_\nu$ grows exponentially as
$a$ goes to infinity, one must set $c_8=0$, and consequently the
first boundary condition~(\ref{bdry}) is satisfied. However, it
is~not easy to find an explicit finite--norm solution to the WD
equation by superposing stationary states, for the integrals over
the order of modified Bessel functions are~not easy to perform.

\item[Case] ${\bf k=-1}$:

The solution for negative curvature is
\begin{equation}
\Psi_E(a, T)=e^{iET}\sqrt{a}\left[c_9J_\mu(6\gamma
a^2)+c_{10}Y_\mu(6\gamma a^2)\right],
\end{equation}
where $c$'s are constants of integrations, $J_\mu$ and $Y_\mu$ are
the Bessel functions and $\mu=\sqrt{1+96\gamma E}/4$. If $0 <
\gamma E < 1/96$ both boundary conditions~(\ref{bdry}) can be
fulfilled, but an explicit wave packet cannot be found by
superposition of stationary states, for very few results are known
for integrals over the order of Bessel functions.
\end{description}

Note that, the above equations for $k=0,\pm 1$ are similar to the
WD equations obtained for perfect fluid cosmological models with
stiff matter in Ref.~\cite{Alvarenga}.

\section{Conclusions and Remarks}

The problem in general theory of relativity goes against the simple Newtonian
picture of the fixed, absolute and external time parameter. The
classical theory, while itself free from problems relating to the
definition and interpretation of time, contains indications of
problems in the quantum theory, where the absence of a time
parameter is hard to reconcile with our everyday experience. In
particular, one of the most fundamental questions in quantum
cosmology, that of identifying a suitable time parameter with
respect to which the dynamics of the Universe can be measured, is unsolved.
Alternatively, in this article, in the Brans--Dicke theory, we address the promising
possibility of the scalar field, that may be originated
from extra dimension in the theory, to resemble the Newtonian external time parameter.
This can be of interest for those physicists
who believe that only an ``external'' time parameter, same as the one
we experience in the Newtonian classical mechanics, can really solve the
problem of time.

In this work,  we first formulate the Brans--Dicke theory of
gravitation in the Einstein frame, in which it is identical to
general relativity with a contribution from a scalar field. Then,
we present the Hamiltonian formulation of spatially homogenous
cosmological model for the theory. Due to the presence of the
minimally coupled scalar field term in the formulation, we show
that the dynamics of the metric functions can be obtained using a
time variable, $T(t)$, as a function of the scalar field. Even
though, the time variable is~not a Dirac observable, it can be
used to label spatial hypersurfaces and plays the role of a
cosmological clock. Besides, the conjugate momentum to this
cosmological clock is a Dirac observable. It has to be emphasized
that the derived dynamical time, with its physical significant, is
emerged from the formulation of the theory in the Einstein frame.
This can be considered as an advantage of this frame over the
Jordan frame in which the Brans--Dicke theory has been written.
Indeed, the introduced cosmological time may justify the Einstein frame with
 the Lagrangian of vacuum general relativity plus a kinetic term of a
scalar field as a physical frame.

We finally apply the results for the classical and quantum FRW
models. We find that the classical models have solutions which
avoid the usual initial cosmological singularity. Particularly, in
a positive curvature space--time, the solution also shows a big
crunch in future while reaches a turning point during its
evolution. In the quantum description of the FRW models, though it
is hard to obtain a physical solution via canonical quantization
due to difficulty in solving the spatial part of the WD equation,
the wave function of universe depends on the dynamical variable
$T$.

\end{document}